\documentclass[                 %
  reprint,
  superscriptaddress,
  altaffilletter,
  amsmath,
  amssymb,
  aps,
  prl,
  floatfix,
]{revtex4-1}

\usepackage{graphicx}
\usepackage{dcolumn}
\usepackage{bm}
\usepackage{soul}
\usepackage{color,hyperref}
\hypersetup{colorlinks,breaklinks,
  linkcolor=blue,urlcolor=blue,
  anchorcolor=blue,citecolor=blue}

\usepackage{multirow}

\usepackage{comment}
\usepackage{lipsum}
\usepackage{mwe}

\newcommand{\onbb}{$0\nu\beta\beta$}
\newcommand{\ky}{kg$\cdot$yr}
\newcommand{\TeOO}{TeO$_2$}

\newcommand{\Te}{$^{130}$Te}
\newcommand{\Tl}{$^{208}$Tl}
\newcommand{\Co}{$^{60}$Co}

\newcommand{\Th}{$^{232}$Th}
\newcommand{\K}{$^{40}$K}
\newcommand{\Qbb}{$Q_{\beta\beta}$}
\newcommand{\rate}{$\Gamma_{0\nu}$}
\newcommand{\bestfitrate}{$\hat{\Gamma}_{0\nu}$}
\newcommand{\rateCo}{$R_{^{60}\text{Co}}$}
\newcommand{\muCo}{$\mu_{^{60}\text{Co}}$}

\newcommand{\mbb}{$m_{\beta\beta}$}
\newcommand{\cuore}{CUORE}
\newcommand{\A}{$\alpha$}

\newcommand{\G}{$\gamma$}

\newcommand{\ckky}{counts$/($keV$\cdot$kg$\cdot$yr$)$}
\newcommand{\cky}{counts$/($kg$\cdot$yr$)$}

\newcommand{\CharacteristicROIEnergyResolutionCalibration}{$7.73(3)$\,keV}
\newcommand{\CharacteristicROIEnergyResolutionBackground}{$7.0(4)$\,keV}
\newcommand{\CharacteristicROIEnergyResolutionBackgroundText}{$7.0\pm0.4$\,keV}

\newcommand{\MaxEnergyBias}{$\leq0.7$\,keV}

\newcommand{\AverageExposurePerMonth}{50}

\newcommand{\ndatasets}{7}
\newcommand{\minNChannels}{900}
\newcommand{\maxNChannels}{954}

\newcommand{\AverageReconstructionEfficiency}{$95.802(3)\,\%$}
\newcommand{\AverageAntiCoincidenceEfficiency}{$98.7(1)\,\%$}
\newcommand{\AveragePSAEfficiency}{$92.6(1)\,\%$}
\newcommand{\AveragePSAEfficiencyText}{$92.6\pm0.1\,\%$}
\newcommand{\SystPSAEfficiencyText}{$\pm0.7\%$}
\newcommand{\AverageTotalAnalEfficiency}{$87.5(2)\,\%$}
\newcommand{\BBContainmentEfficiency}{$88.35(9)\,\%$}

\newcommand{\FinalTotalTeOExposure}{372.5\,\ky}
\newcommand{\FinalCUOREIsotopeExposure}{103.6\,\ky}

\newcommand{\BestFitZeroNuRateDefaultModel}{$(-3.5^{+2.2}_{-1.1})\cdot 10^{-26}$\,yr$^{-1}$}
\newcommand{\CharacteristicROIBackgroundLevel}{$(1.38\pm0.07)\cdot10^{-2}$\,\ckky}
\newcommand{\LowerLimitHalfLifeDefaultMethodStatOnly}{$T^{0\nu}_{1/2}> 3.2\cdot 10^{25}$\,yr}
\newcommand{\LowerLimitHalfLifeDefaultMethodStatOnlyNumberOnly}{$3.2\cdot 10^{25}$\,yr}

\newcommand{\MedianLowerLimitHalfLifeSensitivityDefaultMethod}{$1.7\cdot 10^{25}$\,yr}
\newcommand{\PercProbToGetBetterLimitDefaultMethodStatOnly}{$3\,\%$}

\newcommand{\SystEfficiency}{$0.01\%$}
\newcommand{\SystContEfficiency}{$0.01\%$}
\newcommand{\SystScaling}{$0.02\%$}
\newcommand{\SystQbb}{$0.02\%$}
\newcommand{\SystIsoFrac}{$0.02\%$}
\newcommand{\SystSystEfficiency}{0.04\%}

\newcommand{\LargestSystematic}{$\leq0.04\%$}

\newcommand{\CUOREHalflifeLimit}{\ensuremath{3.2\cdot10^{25}\,{\rm yr}}}
\newcommand{\SystematicOnLimit}{0.4\,\%}

\newcommand{\CombinedMbbLow}{75}
\newcommand{\CombinedMbbHigh}{350\,meV}

\newcommand{\PercChBelowThresholds}{97}

\begin{document}

\title{Improved Limit on Neutrinoless Double-Beta Decay in $^{130}$Te with CUORE}
\author{D.~Q.~Adams}
\affiliation{Department of Physics and Astronomy, University of South Carolina, Columbia, SC 29208, USA}

\author{C.~Alduino}
\affiliation{Department of Physics and Astronomy, University of South Carolina, Columbia, SC 29208, USA}

\author{K.~Alfonso}
\affiliation{Department of Physics and Astronomy, University of California, Los Angeles, CA 90095, USA}

\author{F.~T.~Avignone~III}
\affiliation{Department of Physics and Astronomy, University of South Carolina, Columbia, SC 29208, USA}

\author{O.~Azzolini}
\affiliation{INFN -- Laboratori Nazionali di Legnaro, Legnaro (Padova) I-35020, Italy}

\author{G.~Bari}
\affiliation{INFN -- Sezione di Bologna, Bologna I-40127, Italy}

\author{F.~Bellini}
\affiliation{Dipartimento di Fisica, Sapienza Universit\`{a} di Roma, Roma I-00185, Italy}
\affiliation{INFN -- Sezione di Roma, Roma I-00185, Italy}

\author{G.~Benato}
\affiliation{Department of Physics, University of California, Berkeley, CA 94720, USA}
\affiliation{INFN -- Laboratori Nazionali del Gran Sasso, Assergi (L'Aquila) I-67100, Italy}

\author{M.~Biassoni}
\affiliation{INFN -- Sezione di Milano Bicocca, Milano I-20126, Italy}

\author{A.~Branca}
\affiliation{Dipartimento di Fisica, Universit\`{a} di Milano-Bicocca, Milano I-20126, Italy}
\affiliation{INFN -- Sezione di Milano Bicocca, Milano I-20126, Italy}

% ...

\collaboration{CUORE Collaboration}\noaffiliation
\date{\today}

\begin{abstract}
We report new results from the search for neutrinoless
double-beta decay in \Te\ with the CUORE detector.
This search benefits from a four-fold increase in exposure, lower trigger thresholds
and analysis improvements relative to our previous results.
We observe a background of \CharacteristicROIBackgroundLevel\ in the \onbb\ decay region of interest and,
with a total exposure of \FinalTotalTeOExposure,
we attain a median exclusion sensitivity of \MedianLowerLimitHalfLifeSensitivityDefaultMethod.
We find no evidence for \onbb\ decay  and set a 90\% CI Bayesian lower limit of 
\CUOREHalflifeLimit\ on the \Te\ half-life for this process. 
In the hypothesis that \onbb\ decay is mediated by light Majorana neutrinos,
this results in an upper limit on the effective Majorana mass
of \CombinedMbbLow--\CombinedMbbHigh,
depending on the nuclear matrix elements used.

\end{abstract}


\maketitle

The search for neutrinoless double-beta (\onbb) decay is one of the top
priorities in nuclear and astroparticle physics. 
If observed, this process would unambiguously demonstrate that lepton number is not a
conserved quantity, and the Majorana nature of
neutrinos\,\cite{Racah1937,Furry1939,Pontecorvo:1967fh,BlackboxTheorem}. 
This matter-creating process could provide corroborating 
evidence for the leptogenesis explanation of the baryon asymmetry of the universe\,\cite{Leptogenesis},
and would imply a new mechanism for generating neutrino masses. 
In the simplest scenario whereby \onbb\ decay is mediated by the exchange
of a light Majorana neutrino, the rate of the process depends on the effective Majorana mass \mbb,
though other scenarios exist\,\cite{Minkowski:1977sc,Musolf,Atre:2009rg,Blennow:2010th,Bonnet:2012kh,Mitra:2011qr,Cirigliano17dim7,Cirigliano18master}. 
Even in the absence of direct observation, constraints on the
\onbb\ decay rate can provide important information
on the scale, ordering and origin of neutrino masses\,\cite{Agostini:2017jim}.

Since the \onbb\ decay involves a transition from a nucleus with
$(Z,N)$ protons and neutrons to $(Z+2,N-2)$ with the emission of two
electrons and no neutrinos, the signature is a peak in the summed
energy spectrum of the final-state electrons at the end point of the two neutrino double-beta decay spectrum ($Q_{\beta\beta}$). 
A discovery of the process requires low backgrounds near $Q_{\beta\beta}$,
large masses of isotope, and good energy resolution. 
A wide range of detector technologies is employed in the worldwide
search for \onbb\ decay across a number of isotopes. The current generation
experiments probe half-life values between $10^{25}-10^{26}$\,yr
corresponding to \mbb\ $\mathcal{O}(100$\,meV$)$\,\cite{gerda,majorana,exo,kamlandzen,CUPID-0-final}.
The next generation experiments are planning to instrument
$\mathcal{O}(1000\,\mathrm{kg})$ of isotope and be sensitive to half-lives
beyond $10^{27}$\,yr\,\cite{Agostini:2017jim}.

CUORE\,\cite{CUORE-NIMA,Brofferio:2019yoc} is a ton-scale cryogenic detector
located at Laboratori Nazionali del Gran Sasso (LNGS) in Italy
searching for \onbb\ decay in \Te. 
The experiment consists of an array of 988 \TeOO\ crystals
operating as cryogenic calorimeters\,\cite{Fiorini:1983yj,Enss:2008ek,CUORE-crystals} -- also denoted as bolometers --
at a temperature of about  $10$\,mK.
The detector features excellent energy resolution of $<10$\,keV FWHM in
the \onbb\ region of interest (ROI), large detection efficiency, and low background.
CUORE chose \Te\ because of its high
$Q_{\beta\beta}=(2527.518 \pm 0.013)$\,keV\,\cite{Redshaw:2009cf,Scielzo:2009co,Rahaman:2011wt} --
above most of the natural radioactive background --
and isotopic abundance of $(34.167 \pm 0.002)\%$\,\cite{Fehr:2004jx},
which allows cost-effective use of natural tellurium. 
CUORE is the culmination of decades of development of large-scale bolometric
detectors\,\cite{CUORE-rare-processes,bolometer-saga,CUORE-0-detector,cuoricino-final,CUORE-0-results} and its successful operation
demonstrates the high potential of this technology. 

The CUORE crystals are cubes of $5 \times 5 \times 5$\,cm$^{3}$ and mass of 750\,g\,\cite{CUORE-0-detector},
arranged in 19 towers. 
Each crystal is instrumented with a neutron-transmutation-doped (NTD) germanium  
thermistor\,\cite{Haller1984} to record thermal pulses, and a
silicon heater\,\cite{Alessandrello:1998bf,CUORE-heaters} that
provides reference pulses for thermal gain stabilization. 
The detector is housed in a state-of-the-art
cryostat, which shields the detectors from both thermal and gamma radiation.
Cooling at the 10\,mK stage is achieved by a custom  $^{3}$He/$^{4}$He dilution
refrigerator\,\cite{CUORE-cryostat,CUORE-4K,CUORE-crystals,CUORE-assembly,CUORE-front-end,CUORE-DAQ,CUORE-radon-box,CUORE-PT}.

In this letter, we present an analysis of the data collected between May 2017 and July 2019,
including a reanalysis of data already published\,\cite{CUORE-PRL2017}.
The data are collected in runs of about one day duration,
and grouped into datasets which cover roughly a month.
Each dataset consists of \onbb\ decay search (physics) data bracketed by a few days of calibration
data collected at the beginning and end.
The calibration is performed with either internal \Th\ sources\,\cite{CUORE-dcs} or external mixed \Th-\Co\ sources,
with consecutive datasets sharing the intermediate calibration.
Typically, a few days per dataset are devoted to diagnostics and detector validation measurements,
such as noise optimization\,\cite{CUORE-PT}, working point configuration,
and energy threshold measurements.

\begin{figure}[htbp]
  \includegraphics[width=\columnwidth]{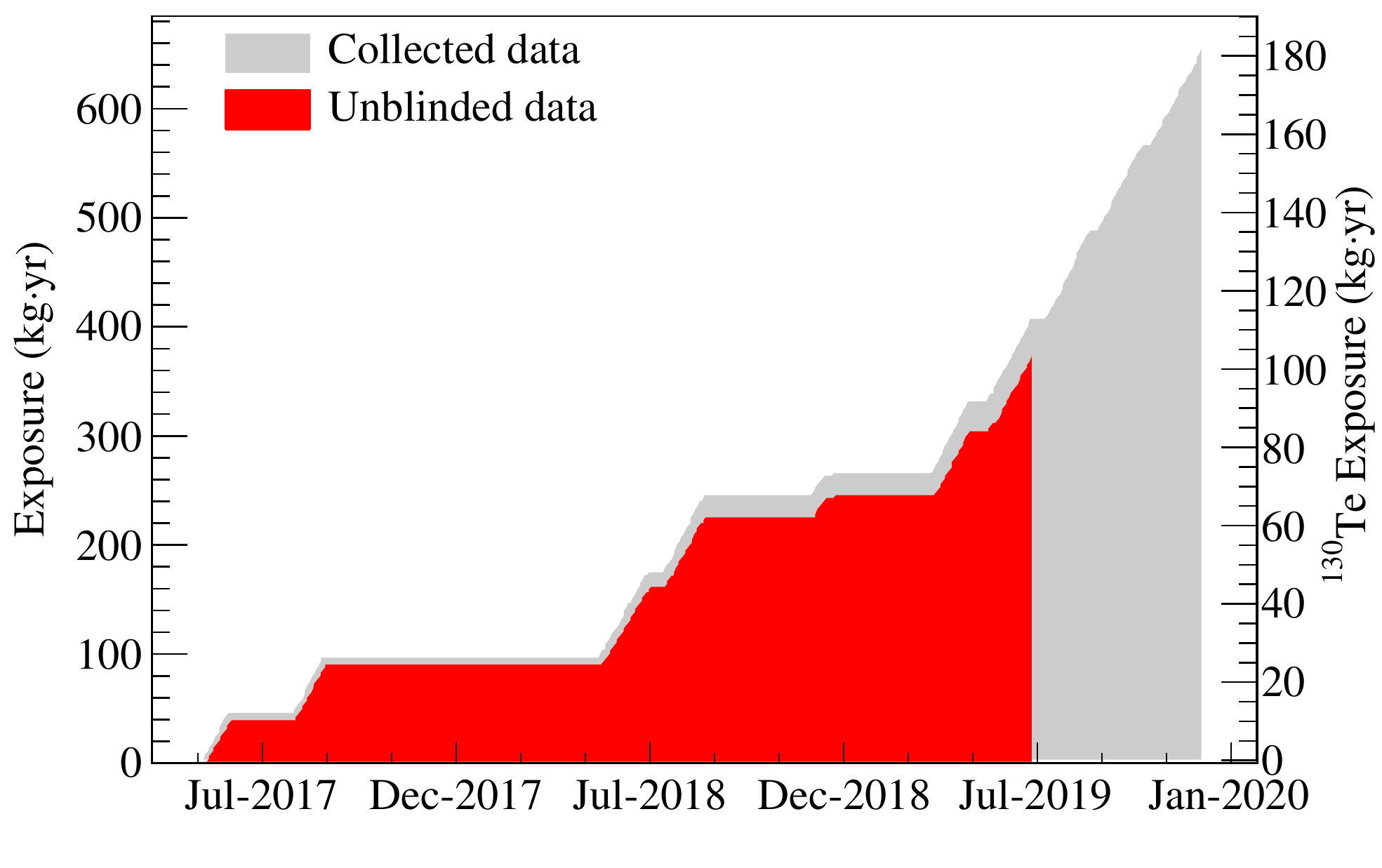}
  \caption{Exposure collected by CUORE starting from May 2017:
    the data in gray represent the exposure accumulated at the time of writing,
    while the data in red are used for the present work (\FinalTotalTeOExposure).}\label{fig:exposure}
\end{figure}

Since we began taking data, there have been two major interruptions of the physics data collection
due to important maintenance of the cryogenic system, as shown in Fig.\,\ref{fig:exposure}.
After the 2017 data campaign\,\cite{CUORE-PRL2017},
we performed a few modifications to improve the stability and uniformity of the data collection:
we decreased the cryostat operating temperature from 15\,mK to 11.8\,mK,
which improves signal-to-noise ratio for injected heater events, 
and installed an external calibration system,
which has a comparable performance to the inner detector calibration system
but is less invasive to deploy.
During the second interruption in fall 2018 we improved the stability of the cryostat
and increased the live-time fraction.
Since spring 2019, CUORE has been stably collecting data
at an average rate of \AverageExposurePerMonth\,\ky$/$month.

\begin{figure}[htbp]
  \includegraphics[width=\columnwidth]{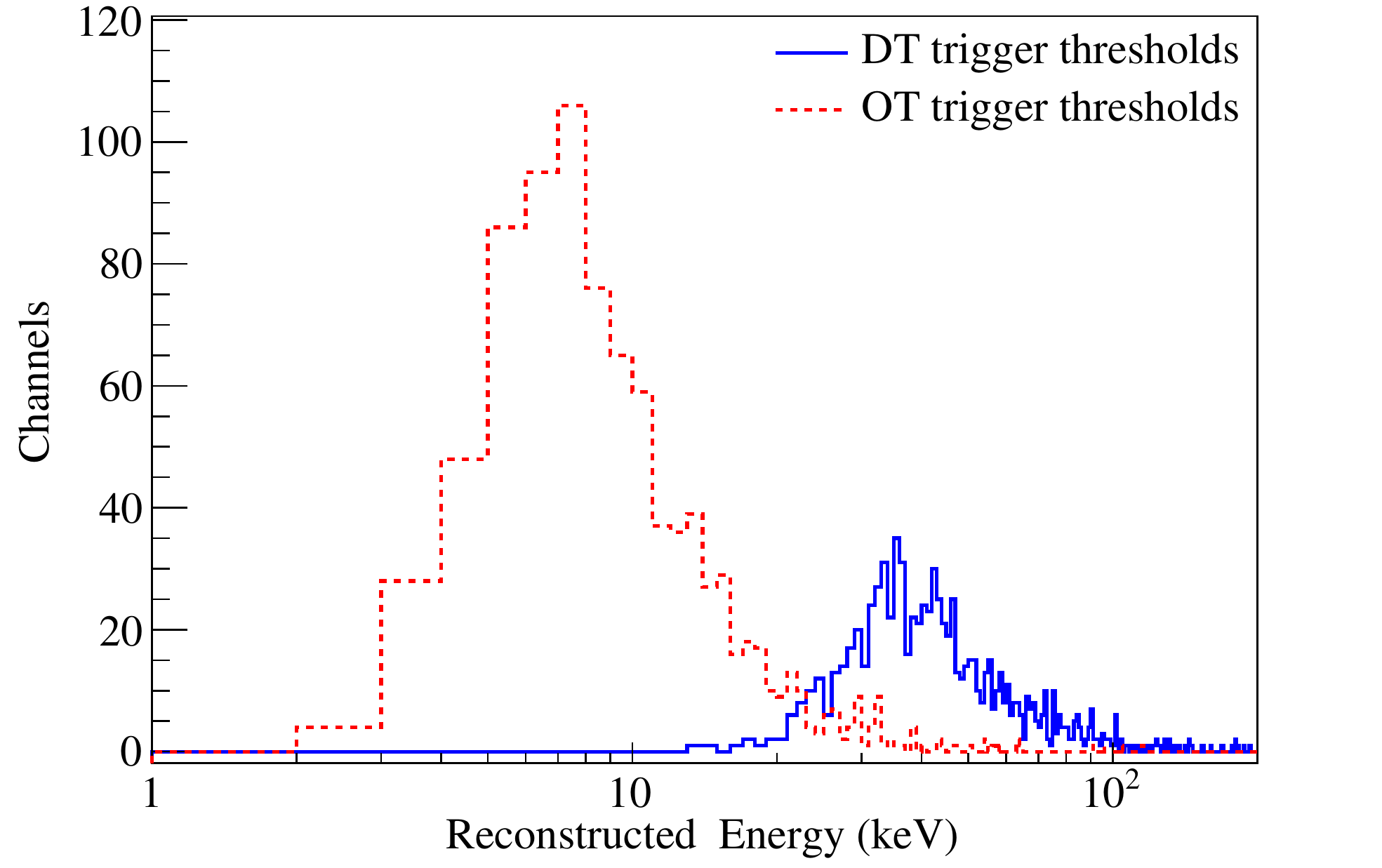}
  \caption{Distribution of energy thresholds at 90\% trigger efficiency for the DT and OT algorithms.}\label{fig:thresholds}
\end{figure}

In this letter, we present an analysis of a \FinalTotalTeOExposure\ \TeOO\ exposure
from the first \ndatasets\ datasets,
corresponding to \FinalCUOREIsotopeExposure\ of \Te\ exposure --
a 4-fold increase in exposure compared to Ref.\,\cite{CUORE-PRL2017}.
The main difference with respect to the first data release is the trigger algorithm.        
The CUORE data acquisition system\,\cite{CUORE-DAQ} saves the full continuous data stream,
giving the possibility of digitally  retriggering the continuous data samples offline.
In this analysis, we use a trigger algorithm based on the optimum filter (OF)\,\cite{Gatti-OT} 
which optimizes the signal-to-noise ratio and yields trigger thresholds
a few times the detector baseline RMS\,\cite{CUORE-0-opttrigger}.
The algorithm identifies a signal if the amplitude of the optimum filtered waveform
exceeds a threshold automatically determined by the baseline resolution of each calorimeter for each dataset.
Figure\,\ref{fig:thresholds} shows the energy thresholds, at 90\% trigger efficiency,
obtained with the previously used algorithm, denoted derivative trigger (DT), and the optimum trigger (OT).
The OT allows us to lower the energy thresholds by a factor 2--10 with respect to the DT.
We set the analysis threshold to 40\,keV to minimize the background contribution in the ROI
from the 2615\,keV \Tl\ line, while at the same time preserving
a $>90\%$ trigger efficiency for the majority ($>$\PercChBelowThresholds\%) of the calorimeters.

The analysis presented here is divided in three parts: event reconstruction,
detector response characterization, and statistical \onbb\ decay analysis.
The goal of the event reconstruction is to extract physical quantities
from each 10\,s long signal waveform.
For each calorimeter and dataset, we build the OF transfer function starting from the average noise
power spectrum and the average signal obtained
from selected calibration events with minimum energies also tuned for each calorimeter and dataset,
and always above 100\,keV.
For each triggered event, we use the waveform filtered with the OF to evaluate the signal amplitude.
To monitor the thermal response of the detector
we use injected heater pulses\,\cite{CUORE-pulser},
and correct the amplitudes of signal events for small temperature drifts.
A second procedure for thermal gain correction  utilizes the \Tl\ events at 2615\,keV from calibration data.
This is the default method for the calorimeters with an unstable pulser,
or not instrumented with a heater.

\begin{figure}[htbp]
  \includegraphics[width=\columnwidth]{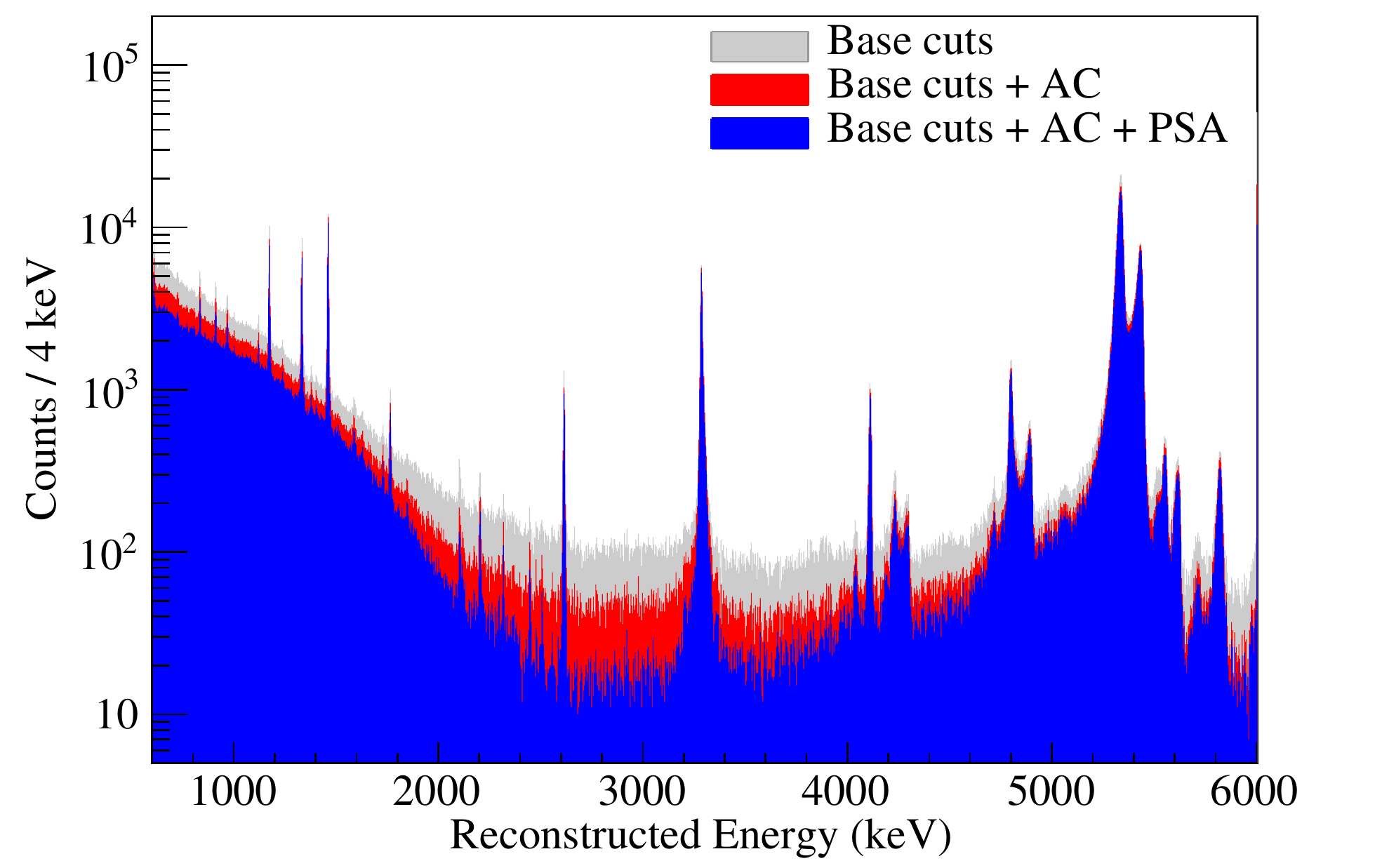}
  \caption{The CUORE spectrum after applying the base cuts to remove heater events
    and periods affected by baseline or noise instabilities (gray),
    after the anticoincidence cut (red), and after the PSA cut (blue).}\label{fig:sumspectrum}
\end{figure}

We use the data acquired in calibration runs to map the stabilized amplitudes to energy values,
and select the stabilization procedure which yields the best energy resolution.
From Monte Carlo (MC) simulations, in $88\%$ of cases, we expect \onbb\ decay events to release energy in just one crystal\,\cite{CUORE-0-analysis-techniques}.
Hence, we apply an anticoincidence cut by computing the number of calorimeters with a $>40$\,keV energy deposition 
within a $\pm5$\,ms time window, and keep only events with a single energy deposition for the \onbb\ decay analysis.
We apply a pulse shape analysis (PSA) to reject events with nonphysical or noisy waveforms,
and pileup -- i.e. superimposed -- events.
We use a set of six pulse shape parameters
to compute the Mahalanobis distance\,\cite{Mahalanobis:1936tj}
from the mean value of a reference sample of clean events from a physical \G\ line.
We tune our cut on the Mahalanobis distance on a per-dataset basis
to optimize the sensitivity to \onbb\ decay events\,\cite{CUORE-PRL2017} (see Fig.\,\ref{fig:sumspectrum}).
The PSA cut mostly affects the continuum regions,
without impacting the \G\  and \A\ lines.
To avoid a human induced bias in the result,
we salt the ROI by moving a random fraction of events
in the $[2615\pm25]$\,keV region into the $[$\Qbb$\pm25]$\,keV region,
and vice versa\,\cite{CUORE-PRL2017,CUORE-0-analysis-techniques}.
The salting is reversed once the analysis procedures are finalized.

The signal efficiency is calculated as the product of the containment efficiency,
the trigger and reconstruction efficiency, the anticoincidence efficiency, and the PSA efficiency.
We compute the probability that the full energy of \onbb\ decay is contained in a single crystal
using MC simulations.
We use the injected heater pulses to evaluate the efficiency of correctly triggering all injected events, 
reconstructing their pulse energy\,\cite{CUORE-PRL2017},
as well as the probability of false positives in the identification
of pileup events.
Given the large number of heater events, the trigger, energy reconstruction
and pileup rejection efficiencies are obtained for each calorimeter and dataset separately,
then averaged over the entire dataset.

We extract the anticoincidence efficiency
as the survival probability of the 1460\,keV \G\ line from \K\ in the physics data.
This \G\ line follows an electron capture with a K-shell energy of $\sim3$\,keV,
well below threshold; thus a fully absorbed 1460\,keV \K\ \G\ line is
uncorrelated to any other event.
Finally, we compute the PSA efficiency by calculating the survival probability
of two samples of physical events:
double-crystal events whose sum energy is compatible with \G\ lines,
or single crystal events corresponding to fully absorbed \G\ lines.
The former method utilizes events at all energies, since only the summed energy is fixed,
and provides a cleaner but smaller data sample.
The latter method profits from higher statistics for most \G\ lines,
but it requires background substraction and
only allows for an efficiency determination at a handful of energies.
We choose the PSA efficiency 
as the average of the efficiencies obtained from these two samples (\AveragePSAEfficiencyText)
and treat the difference between them as a systematic effect, 
adding a scaling parameter common to all datasets in the final fit (\SystPSAEfficiencyText). 
Given the limited statistics of the physics data, the anticoincidence and PSA efficiencies
can only be extracted for an entire dataset, and have larger uncertainties
than the efficiencies obtained from heater data.
The exposure-weighted average efficiencies are reported in Tab.\,\ref{tab:parameters}.

We extract the detector response function in the ROI for each calorimeter in each dataset
by fitting the 2615\,keV \Tl\ line  in the calibration spectrum\,\cite{alexey,laura}.
To evaluate possible systematic shifts in the energy scale
and the energy dependence of the detector energy resolution,
we use the detector response function obtained from the \Tl\ calibration peak,
with the addition of a linear function to model the background,
to fit the 5--7 most prominent $\gamma$ lines of the physics spectrum.
We keep as free parameters the peak position, the peak amplitude,
and the ratio of the energy resolution in physics and calibration data.
We extract the energy calibration bias
-- defined as the difference between the reconstructed peak position and its nominal value --
and energy resolution, parameterize them quadratically as a function of energy,
and interpolate them to \Qbb.
The exposure-weighted harmonic average of the energy resolution at \Qbb\ in the physics data
is \CharacteristicROIEnergyResolutionBackgroundText, 
while the energy bias is \MaxEnergyBias.
A summary of relevant quantities for the \onbb\ decay analysis is given in Tab.\,\ref{tab:parameters}.

\begin{table}[htbp]
  \caption{Relevant quantities and effective parameters of the analysis. The FWHM of calibration data
    is the exposure weighted harmonic average over all calorimeters and datasets,
    which is projected to \Qbb\ in the physics data. The containment efficiency
    is from MC simulations, while all other efficiencies correspond
    to the exposure weighted means.}\label{tab:parameters}
  \begin{tabular}{rcl}
    \toprule
    Number of datasets                    &  & \ndatasets \\
    Number of valid calorimeters (min--max)         &  & \minNChannels --\maxNChannels \\
    \TeOO\ exposure                       &  & \FinalTotalTeOExposure \\
    FWHM at 2615\,keV in calibration data &  & \CharacteristicROIEnergyResolutionCalibration \\
    FWHM at \Qbb\ in physics data         &  & \CharacteristicROIEnergyResolutionBackground \\
    Reconstruction efficiency             &  & \AverageReconstructionEfficiency \\
    Anticoincidence efficiency           &  & \AverageAntiCoincidenceEfficiency \\
    PSA efficiency                        &  & \AveragePSAEfficiency \\
    Total analysis efficiency             &  & \AverageTotalAnalEfficiency \\
    Containment efficiency                &  & \BBContainmentEfficiency\,\cite{CUORE-0-analysis-techniques} \\
    \botrule
  \end{tabular}
\end{table}

\begin{figure}[htbp]
  \includegraphics[width=\columnwidth]{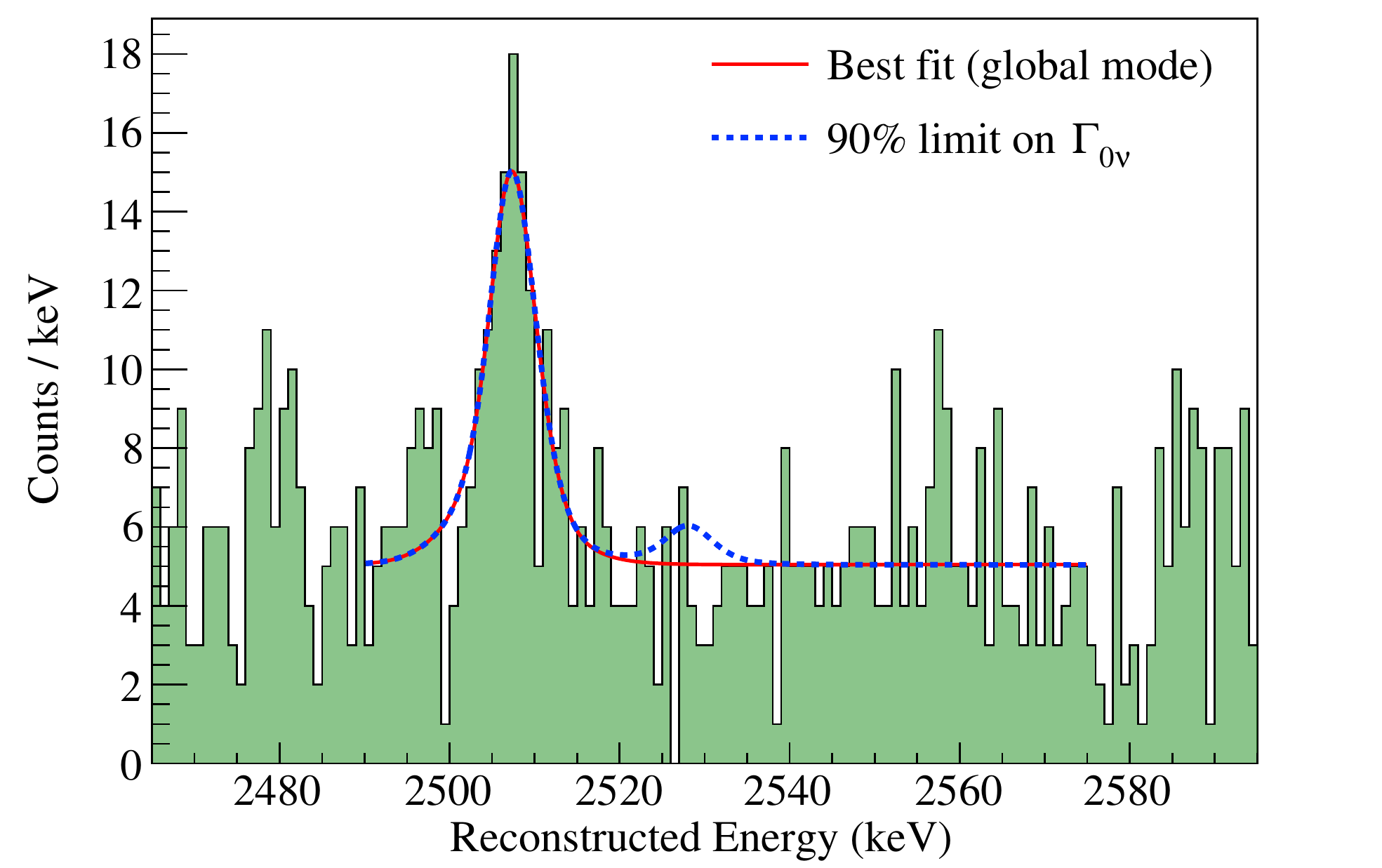}
  \caption{ROI spectrum with the best-fit curve (solid red) and
    the best fit-curve with the \onbb\ decay component fixed to the 90\% CI limit (dashed blue).}\label{fig:bestfit}
\end{figure}

The \cuore\ physics spectrum (Fig.\,\ref{fig:bestfit}) around \Qbb\ features a flat distribution
with $\sim90\%$ of the events coming from degraded \A\ particles,
as obtained by extrapolating from the flat \A\ background
in the energy region above the 2615\,keV \Tl\ line,
and $\sim10\%$ from 2615\,keV \G\ events
undergoing multiple Compton scattering~\cite{CUORE-projected-background,CUORE-0-bkgmodel}.
The closest expected peak to \Qbb\ is the \Co\ sum peak at 2505.7\,keV.
We find an additional structure
with a significance of $\gtrsim$2\,$\sigma$ at $\sim$2480\,keV,
visible only in the single-crystal spectrum.
Its energy corresponds to a \Co\ sum peak, with an escaping Te xray,
but its amplitude is much larger than expected from MC simulations,
and it is not visible in \Co\ calibration spectra.
We considered various possible contamination, 
but none justifies the presence of a peak at $\sim$2480\,keV with the observed rate.
Thus, more data are needed to assess the significance of this feature.
As a consequence, we restrict the fit range to [2490,2575]\,keV region,
and fit the data with a flat background plus peaks described by the
detector response function
for  the \Co\ sum line and the potential \onbb\ decay signal.

We perform an unbinned Bayesian fit combined over all datasets using the BAT software package\,\cite{bat}.
The model parameters are the \onbb\ decay rate (\rate),
a dataset dependent background index (BI) in \ckky,
the \Co\ sum peak amplitude (\rateCo) in \cky, and its position \muCo,
which is a free parameter as in the previous analysis\,\cite{CUORE-PRL2017}.
The BIs are dataset dependent, while all other parameters are common to all datasets,
including the \Co\ rate, which is scaled by a dataset dependent factor to account for its decay.
We use flat priors for all of the parameters, and restrict the range of the BIs
and all peak rates to the physical range, i.e. non-negative values.

\begin{figure}[htbp]
  \includegraphics[width=\columnwidth]{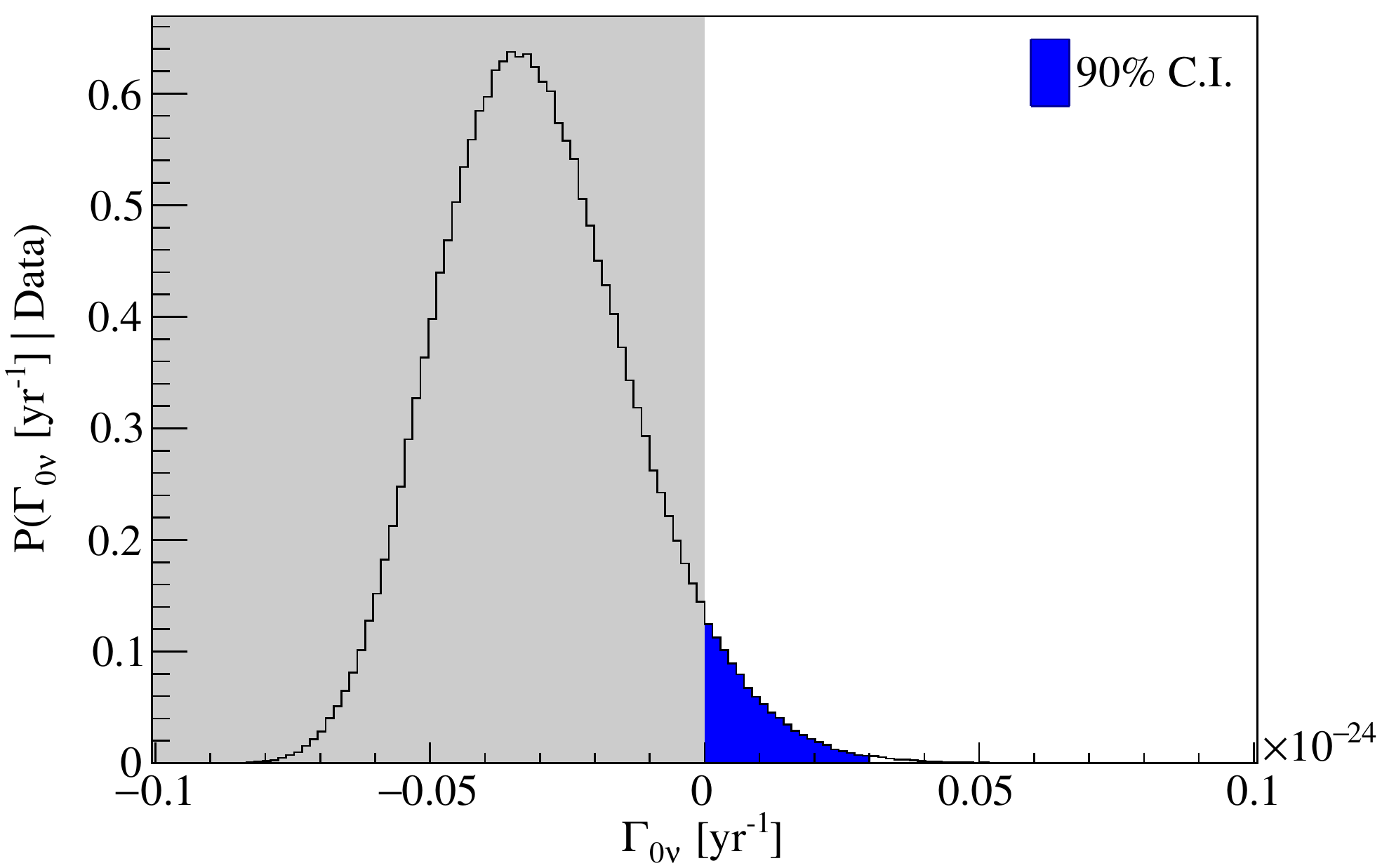}
  \caption{Posterior on \rate\ with all systematics included
    for the fit on the physical range (\rate$>0$) and on the full range.
    The 90\% CI is shown in blue.}\label{fig:posterior}
\end{figure}

We find no evidence for \onbb\ decay,
obtaining the \rate\ posterior distribution reported in Fig.\,\ref{fig:posterior},
and a limit of \LowerLimitHalfLifeDefaultMethodStatOnly\ at 90\% credibility interval (CI),
corresponding to the blue dashed curve in Fig.\,\ref{fig:bestfit}.
Repeating the fit without the \onbb\ decay contribution,
we obtain an average BI of \CharacteristicROIBackgroundLevel.

To compute the exclusion sensitivity, we generate $10^4$ sets of pseudo-experiments
populated with only the \Co\ and flat background components,
and divided into 7 datasets with the same exposure and BI of the actual datasets
as obtained from fitting the data with the background-only model.
We fit each pseudo-experiment with the standard signal-plus-background model
and obtain a median 90\% CI exclusion sensitivity of \MedianLowerLimitHalfLifeSensitivityDefaultMethod.
The probability of obtaining a stronger limit than \LowerLimitHalfLifeDefaultMethodStatOnlyNumberOnly\
is \PercProbToGetBetterLimitDefaultMethodStatOnly.

We consider the following systematic effects.
The dominant one is the systematic error on the PSA efficiency.
Subdominant effects are induced by the uncertainties
on the energy scale, energy resolution, analysis and containment efficiencies, the value of \Qbb,
and the \Te\ natural isotopic abundance.
We implement all systematics as additional nuisance parameters in the fit,
which can be activated independently,
with the priors reported in Tab.\,\ref{tab:systematics}.
We evaluate the bias induced by each nuisance parameter by looking at the effect on \rate\
at the posterior global mode, \bestfitrate; for this, we artificially release the \rate$\geq0$ constraint,
allowing \rate\ to be negative.
The best fit with the artificially extended range
is \bestfitrate=\BestFitZeroNuRateDefaultModel, with a $\sim1.6\,\sigma$ background under-fluctuation
compatible with the p-value quoted above.
The inclusion of the additional nuisance parameters affects the global mode on \rate\ by \LargestSystematic\ (see also Tab.\,\ref{tab:systematics}),
and yields a \SystematicOnLimit\ weaker limit.

Assuming \onbb\ decay is mediated by light neutrino exchange,
and using the phase space factor from Ref.~\cite{Kotila:2012zza},
the result above corresponds to a set of upper limits on the effective Majorana mass
ranging between \CombinedMbbLow\ and \CombinedMbbHigh,
where the spread reflects the different nuclear matrix element calculations
available in literature\,\cite{Engel:2016xgb,Barea:2015kwa,Simkovic:2013qiy,Hyvarinen:2015bda,Menendez:2008jp,Rodriguez:2010mn,Vaquero:2014dna,Yao:2014uta,Mustonen:2013zu,Neacsu:2014bia,Meroni:2012qf}.
The results reported in this paper represent the most stringent limit on \onbb\ decay in \Te,
and our limit on \mbb\ is competitive with the leading ones in the field\,\cite{gerda,majorana,exo,kamlandzen,CUPID-0-final}.

\begin{table}[htbp]
  \caption{Systematics affecting the \onbb\ decay analysis. Total analysis efficiency I corresponds
    to the product of all efficiency terms reported in Tab\,\ref{tab:parameters},
    while total analysis efficiency II corresponds to the additional systematic on the PSA efficiency.
    We report the systematic on the \rate\ global mode obtained
    leaving \rate\ free to assume negative (nonphysical) values.}\label{tab:systematics}
  \begin{tabular}{rlc}
    \toprule
    \multicolumn{3}{c}{Fit parameter systematics} \\
    \colrule
    Systematic                     & Prior        & Effect on \bestfitrate  \\
    \colrule
    Total analysis efficiency I          & Gaussian     & \SystEfficiency  \\
    Total analysis efficiency II         & Uniform      & \SystSystEfficiency  \\
    Containment efficiency         & Gaussian     & \SystContEfficiency \\
    Energy and resolution scaling  & Multivariate & \SystScaling \\
    \Qbb                           & Gaussian     & \SystQbb \\
    Isotopic fraction              & Gaussian     & \SystIsoFrac \\
    \botrule
  \end{tabular}
\end{table}

After a period of detector maintenance and optimization,
CUORE is stably collecting data at a rate of \AverageExposurePerMonth\,\ky$/$month.
The experiment is a proof of the power and scalability of the bolometric technique to the ton scale.
Recent developments in scintillating crystals demonstrate the technology
for a future zero-background search\,\cite{CUORE-projected-background,CUORE-0-bkgmodel,CUPID-0-final,CUPID-Mo-instrument,CUPID-preCDR}.
These advances will be exploited in the next generation bolometric experiment,
CUPID\,\cite{CUPID-preCDR}, which will reuse the CUORE cryostat and infrastructure.


\section{Acknowledgments}

The CUORE Collaboration thanks the directors and staff
of the Laboratori Nazionali del Gran Sasso and the technical staff of our laboratories.
This work was supported by the Istituto Nazionale di Fisica Nucleare (INFN);
the National Science Foundation under Grant Nos. NSF-PHY-0605119, NSF-PHY-0500337, NSF-PHY-0855314,
NSF-PHY-0902171, NSF-PHY-0969852, NSF-PHY-1307204, NSF-PHY-1314881, NSF-PHY-1401832, and NSF-PHY-1404205;
the Alfred P. Sloan Foundation; the University of Wisconsin Foundation; and Yale University.
This material is also based upon work supported by the US Department of Energy (DOE) Office of Science under Contract Nos.
DE-AC02-05CH11231, DE-AC52-07NA27344, DE-SC0012654, and  DE-SC0020423;
by the DOE Office of Science, Office of Nuclear Physics under Contract Nos. DE-FG02-08ER41551 and DE-FG03-00ER41138;
and by the European Union’s Horizon 2020 research and innovation program
under the Marie Sklodowska-Curie grant agreement No 754496.
This research used resources of the National Energy Research Scientific Computing Center (NERSC).
This work makes use of the DIANA data analysis and APOLLO data acquisition software which has been developed
by the CUORICINO, CUORE, LUCIFER and CUPID-0 collaborations.

\bibliography{Bibliography} 

\begin{thebibliography}{69}%
\makeatletter
\providecommand \@ifxundefined [1]{%
 \@ifx{#1\undefined}
}%
\providecommand \@ifnum [1]{%
 \ifnum #1\expandafter \@firstoftwo
 \else \expandafter \@secondoftwo
 \fi
}%
\providecommand \@ifx [1]{%
 \ifx #1\expandafter \@firstoftwo
 \else \expandafter \@secondoftwo
 \fi
}%
\providecommand \natexlab [1]{#1}%
\providecommand \enquote  [1]{``#1''}%
\providecommand \bibnamefont  [1]{#1}%
\providecommand \bibfnamefont [1]{#1}%
\providecommand \citenamefont [1]{#1}%
\providecommand \href@noop [0]{\@secondoftwo}%
\providecommand \href [0]{\begingroup \@sanitize@url \@href}%
\providecommand \@href[1]{\@@startlink{#1}\@@href}%
\providecommand \@@href[1]{\endgroup#1\@@endlink}%
\providecommand \@sanitize@url [0]{\catcode `\\12\catcode `\$12\catcode
  `\&12\catcode `\#12\catcode `\^12\catcode `\_12\catcode `\%12\relax}%
\providecommand \@@startlink[1]{}%
\providecommand \@@endlink[0]{}%
\providecommand \url  [0]{\begingroup\@sanitize@url \@url }%
\providecommand \@url [1]{\endgroup\@href {#1}{\urlprefix }}%
\providecommand \urlprefix  [0]{URL }%
\providecommand \Eprint [0]{\href }%
\providecommand \doibase [0]{http://dx.doi.org/}%
\providecommand \selectlanguage [0]{\@gobble}%
\providecommand \bibinfo  [0]{\@secondoftwo}%
\providecommand \bibfield  [0]{\@secondoftwo}%
\providecommand \translation [1]{[#1]}%
\providecommand \BibitemOpen [0]{}%
\providecommand \bibitemStop [0]{}%
\providecommand \bibitemNoStop [0]{.\EOS\space}%
\providecommand \EOS [0]{\spacefactor3000\relax}%
\providecommand \BibitemShut  [1]{\csname bibitem#1\endcsname}%
\let\auto@bib@innerbib\@empty
\bibitem [{\citenamefont {Racah}(1937)}]{Racah1937}%
  \BibitemOpen
  \bibfield  {author} {\bibinfo {author} {\bibfnamefont {G.}~\bibnamefont
  {Racah}},\ }\href {\doibase 10.1007/BF02961321} {\bibfield  {journal}
  {\bibinfo  {journal} {Nuovo Cimento}\ }\textbf {\bibinfo {volume} {14}},\
  \bibinfo {pages} {322} (\bibinfo {year} {1937})}\BibitemShut {NoStop}%
\bibitem [{\citenamefont {Furry}(1939)}]{Furry1939}%
  \BibitemOpen
  \bibfield  {author} {\bibinfo {author} {\bibfnamefont {W.~H.}\ \bibnamefont
  {Furry}},\ }\href {\doibase 10.1103/PhysRev.56.1184} {\bibfield  {journal}
  {\bibinfo  {journal} {Phys. Rev.}\ }\textbf {\bibinfo {volume} {56}},\
  \bibinfo {pages} {1184} (\bibinfo {year} {1939})}\BibitemShut {NoStop}%
\bibitem [{\citenamefont {Pontecorvo}(1968)}]{Pontecorvo:1967fh}%
  \BibitemOpen
  \bibfield  {author} {\bibinfo {author} {\bibfnamefont {B.}~\bibnamefont
  {Pontecorvo}},\ }\href@noop {} {\bibfield  {journal} {\bibinfo  {journal}
  {Sov. Phys. JETP}\ }\textbf {\bibinfo {volume} {26}},\ \bibinfo {pages} {984}
  (\bibinfo {year} {1968})}\BibitemShut {NoStop}%
\bibitem [{\citenamefont {Schechter}\ and\ \citenamefont
  {Valle}(1982)}]{BlackboxTheorem}%
  \BibitemOpen
  \bibfield  {author} {\bibinfo {author} {\bibfnamefont {J.}~\bibnamefont
  {Schechter}}\ and\ \bibinfo {author} {\bibfnamefont {J.~W.~F.}\ \bibnamefont
  {Valle}},\ }\href {\doibase 10.1103/PhysRevD.25.2951} {\bibfield  {journal}
  {\bibinfo  {journal} {Phys. Rev.}\ }\textbf {\bibinfo {volume} {D25}},\
  \bibinfo {pages} {2951} (\bibinfo {year} {1982})}\BibitemShut {NoStop}%
\bibitem [{\citenamefont {Fukugita}\ and\ \citenamefont
  {Yanagida}(1986)}]{Leptogenesis}%
  \BibitemOpen
  \bibfield  {author} {\bibinfo {author} {\bibfnamefont {M.}~\bibnamefont
  {Fukugita}}\ and\ \bibinfo {author} {\bibfnamefont {T.}~\bibnamefont
  {Yanagida}},\ }\href {\doibase https://doi.org/10.1016/0370-2693(86)91126-3}
  {\bibfield  {journal} {\bibinfo  {journal} {Phys. Lett. B}\ }\textbf
  {\bibinfo {volume} {174}},\ \bibinfo {pages} {45 } (\bibinfo {year}
  {1986})}\BibitemShut {NoStop}%
\bibitem [{\citenamefont {Minkowski}(1977)}]{Minkowski:1977sc}%
  \BibitemOpen
  \bibfield  {author} {\bibinfo {author} {\bibfnamefont {P.}~\bibnamefont
  {Minkowski}},\ }\href {\doibase 10.1016/0370-2693(77)90435-X} {\bibfield
  {journal} {\bibinfo  {journal} {Phys. Lett.}\ }\textbf {\bibinfo {volume}
  {67B}},\ \bibinfo {pages} {421} (\bibinfo {year} {1977})}\BibitemShut
  {NoStop}%
\bibitem [{\citenamefont {Prezeau}\ \emph {et~al.}(2003)\citenamefont
  {Prezeau}, \citenamefont {Ramsey-Musolf},\ and\ \citenamefont
  {Vogel}}]{Musolf}%
  \BibitemOpen
  \bibfield  {author} {\bibinfo {author} {\bibfnamefont {G.}~\bibnamefont
  {Prezeau}}, \bibinfo {author} {\bibfnamefont {M.}~\bibnamefont
  {Ramsey-Musolf}}, \ and\ \bibinfo {author} {\bibfnamefont {P.}~\bibnamefont
  {Vogel}},\ }\href {\doibase 10.1103/PhysRevD.68.034016} {\bibfield  {journal}
  {\bibinfo  {journal} {Phys. Rev.}\ }\textbf {\bibinfo {volume} {D68}},\
  \bibinfo {pages} {034016} (\bibinfo {year} {2003})}\BibitemShut {NoStop}%
\bibitem [{\citenamefont {Atre}\ \emph {et~al.}(2009)\citenamefont {Atre} \emph
  {et~al.}}]{Atre:2009rg}%
  \BibitemOpen
  \bibfield  {author} {\bibinfo {author} {\bibfnamefont {A.}~\bibnamefont
  {Atre}} \emph {et~al.},\ }\href {\doibase 10.1088/1126-6708/2009/05/030}
  {\bibfield  {journal} {\bibinfo  {journal} {JHEP}\ }\textbf {\bibinfo
  {volume} {05}},\ \bibinfo {pages} {030} (\bibinfo {year} {2009})}\BibitemShut
  {NoStop}%
\bibitem [{\citenamefont {Blennow}\ \emph {et~al.}(2010)\citenamefont {Blennow}
  \emph {et~al.}}]{Blennow:2010th}%
  \BibitemOpen
  \bibfield  {author} {\bibinfo {author} {\bibfnamefont {M.}~\bibnamefont
  {Blennow}} \emph {et~al.},\ }\href {\doibase 10.1007/JHEP07(2010)096}
  {\bibfield  {journal} {\bibinfo  {journal} {JHEP}\ }\textbf {\bibinfo
  {volume} {07}},\ \bibinfo {pages} {096} (\bibinfo {year} {2010})}\BibitemShut
  {NoStop}%
\bibitem [{\citenamefont {Bonnet}\ \emph {et~al.}(2013)\citenamefont {Bonnet}
  \emph {et~al.}}]{Bonnet:2012kh}%
  \BibitemOpen
  \bibfield  {author} {\bibinfo {author} {\bibfnamefont {F.}~\bibnamefont
  {Bonnet}} \emph {et~al.},\ }\href {\doibase 10.1007/JHEP03(2013)055}
  {\bibfield  {journal} {\bibinfo  {journal} {JHEP}\ }\textbf {\bibinfo
  {volume} {03}},\ \bibinfo {pages} {055} (\bibinfo {year} {2013})},\ \bibinfo
  {note} {[Erratum: JHEP04,090(2014)]}\BibitemShut {NoStop}%
\bibitem [{\citenamefont {Mitra}\ \emph {et~al.}(2012)\citenamefont {Mitra},
  \citenamefont {Senjanovic},\ and\ \citenamefont {Vissani}}]{Mitra:2011qr}%
  \BibitemOpen
  \bibfield  {author} {\bibinfo {author} {\bibfnamefont {M.}~\bibnamefont
  {Mitra}}, \bibinfo {author} {\bibfnamefont {G.}~\bibnamefont {Senjanovic}}, \
  and\ \bibinfo {author} {\bibfnamefont {F.}~\bibnamefont {Vissani}},\ }\href
  {\doibase 10.1016/j.nuclphysb.2011.10.035} {\bibfield  {journal} {\bibinfo
  {journal} {Nucl. Phys.}\ }\textbf {\bibinfo {volume} {B856}},\ \bibinfo
  {pages} {26} (\bibinfo {year} {2012})}\BibitemShut {NoStop}%
\bibitem [{\citenamefont {Cirigliano}\ \emph {et~al.}(2017)\citenamefont
  {Cirigliano} \emph {et~al.}}]{Cirigliano17dim7}%
  \BibitemOpen
  \bibfield  {author} {\bibinfo {author} {\bibfnamefont {V.}~\bibnamefont
  {Cirigliano}} \emph {et~al.},\ }\href {\doibase 10.1007/JHEP12(2017)082}
  {\bibfield  {journal} {\bibinfo  {journal} {JHEP}\ }\textbf {\bibinfo
  {volume} {12}},\ \bibinfo {pages} {082} (\bibinfo {year} {2017})}\BibitemShut
  {NoStop}%
\bibitem [{\citenamefont {Cirigliano}\ \emph {et~al.}(2018)\citenamefont
  {Cirigliano} \emph {et~al.}}]{Cirigliano18master}%
  \BibitemOpen
  \bibfield  {author} {\bibinfo {author} {\bibfnamefont {V.}~\bibnamefont
  {Cirigliano}} \emph {et~al.},\ }\href {\doibase 10.1007/JHEP12(2018)097}
  {\bibfield  {journal} {\bibinfo  {journal} {JHEP}\ }\textbf {\bibinfo
  {volume} {12}},\ \bibinfo {pages} {097} (\bibinfo {year} {2018})}\BibitemShut
  {NoStop}%
\bibitem [{\citenamefont {Agostini}\ \emph {et~al.}(2017)\citenamefont
  {Agostini}, \citenamefont {Benato},\ and\ \citenamefont
  {Detwiler}}]{Agostini:2017jim}%
  \BibitemOpen
  \bibfield  {author} {\bibinfo {author} {\bibfnamefont {M.}~\bibnamefont
  {Agostini}}, \bibinfo {author} {\bibfnamefont {G.}~\bibnamefont {Benato}}, \
  and\ \bibinfo {author} {\bibfnamefont {J.}~\bibnamefont {Detwiler}},\ }\href
  {\doibase 10.1103/PhysRevD.96.053001} {\bibfield  {journal} {\bibinfo
  {journal} {Phys. Rev.}\ }\textbf {\bibinfo {volume} {D96}},\ \bibinfo {pages}
  {053001} (\bibinfo {year} {2017})}\BibitemShut {NoStop}%
\bibitem [{\citenamefont {Agostini}\ \emph {et~al.}(2019)\citenamefont
  {Agostini} \emph {et~al.}}]{gerda}%
  \BibitemOpen
  \bibfield  {author} {\bibinfo {author} {\bibfnamefont {M.}~\bibnamefont
  {Agostini}} \emph {et~al.} (\bibinfo {collaboration} {GERDA}),\ }\href
  {\doibase 10.1126/science.aav8613} {\bibfield  {journal} {\bibinfo  {journal}
  {Science}\ }\textbf {\bibinfo {volume} {365}},\ \bibinfo {pages} {1445}
  (\bibinfo {year} {2019})}\BibitemShut {NoStop}%
\bibitem [{\citenamefont {Alvis}\ \emph {et~al.}(2019)\citenamefont {Alvis}
  \emph {et~al.}}]{majorana}%
  \BibitemOpen
  \bibfield  {author} {\bibinfo {author} {\bibfnamefont {S.~I.}\ \bibnamefont
  {Alvis}} \emph {et~al.} (\bibinfo {collaboration} {Majorana}),\ }\href
  {\doibase 10.1103/PhysRevC.100.025501} {\bibfield  {journal} {\bibinfo
  {journal} {Phys. Rev.}\ }\textbf {\bibinfo {volume} {C100}},\ \bibinfo
  {pages} {025501} (\bibinfo {year} {2019})}\BibitemShut {NoStop}%
\bibitem [{\citenamefont {Anton}\ \emph {et~al.}(2019)\citenamefont {Anton}
  \emph {et~al.}}]{exo}%
  \BibitemOpen
  \bibfield  {author} {\bibinfo {author} {\bibfnamefont {G.}~\bibnamefont
  {Anton}} \emph {et~al.} (\bibinfo {collaboration} {EXO-200}),\ }\href
  {\doibase 10.1103/PhysRevLett.123.161802} {\bibfield  {journal} {\bibinfo
  {journal} {Phys. Rev. Lett.}\ }\textbf {\bibinfo {volume} {123}},\ \bibinfo
  {pages} {161802} (\bibinfo {year} {2019})}\BibitemShut {NoStop}%
\bibitem [{\citenamefont {Gando}\ \emph {et~al.}(2016)\citenamefont {Gando}
  \emph {et~al.}}]{kamlandzen}%
  \BibitemOpen
  \bibfield  {author} {\bibinfo {author} {\bibfnamefont {A.}~\bibnamefont
  {Gando}} \emph {et~al.} (\bibinfo {collaboration} {KamLAND-Zen}),\ }\href
  {\doibase 10.1103/PhysRevLett.117.109903, 10.1103/PhysRevLett.117.082503}
  {\bibfield  {journal} {\bibinfo  {journal} {Phys. Rev. Lett.}\ }\textbf
  {\bibinfo {volume} {117}},\ \bibinfo {pages} {082503} (\bibinfo {year}
  {2016})},\ \bibinfo {note} {[Addendum: Phys. Rev.
  Lett.117,no.10,109903(2016)]}\BibitemShut {NoStop}%
\bibitem [{\citenamefont {Azzolini}\ \emph {et~al.}(2019)\citenamefont
  {Azzolini} \emph {et~al.}}]{CUPID-0-final}%
  \BibitemOpen
  \bibfield  {author} {\bibinfo {author} {\bibfnamefont {O.}~\bibnamefont
  {Azzolini}} \emph {et~al.} (\bibinfo {collaboration} {CUPID-0}),\ }\href
  {\doibase 10.1103/PhysRevLett.123.032501} {\bibfield  {journal} {\bibinfo
  {journal} {Phys. Rev. Lett.}\ }\textbf {\bibinfo {volume} {123}},\ \bibinfo
  {pages} {032501} (\bibinfo {year} {2019})}\BibitemShut {NoStop}%
\bibitem [{\citenamefont {Arnaboldi}\ \emph {et~al.}(2004)\citenamefont
  {Arnaboldi} \emph {et~al.}}]{CUORE-NIMA}%
  \BibitemOpen
  \bibfield  {author} {\bibinfo {author} {\bibfnamefont {C.}~\bibnamefont
  {Arnaboldi}} \emph {et~al.} (\bibinfo {collaboration} {CUORE}),\ }\href
  {\doibase 10.1016/j.nima.2003.07.067} {\bibfield  {journal} {\bibinfo
  {journal} {Nucl. Instrum. Meth.}\ }\textbf {\bibinfo {volume} {A518}},\
  \bibinfo {pages} {775} (\bibinfo {year} {2004})}\BibitemShut {NoStop}%
\bibitem [{\citenamefont {Brofferio}\ \emph {et~al.}(2019)\citenamefont
  {Brofferio}, \citenamefont {Cremonesi},\ and\ \citenamefont
  {Dell'Oro}}]{Brofferio:2019yoc}%
  \BibitemOpen
  \bibfield  {author} {\bibinfo {author} {\bibfnamefont {C.}~\bibnamefont
  {Brofferio}}, \bibinfo {author} {\bibfnamefont {O.}~\bibnamefont
  {Cremonesi}}, \ and\ \bibinfo {author} {\bibfnamefont {S.}~\bibnamefont
  {Dell'Oro}},\ }\href {\doibase 10.3389/fphy.2019.00086} {\bibfield  {journal}
  {\bibinfo  {journal} {Front.in Phys.}\ }\textbf {\bibinfo {volume} {7}},\
  \bibinfo {pages} {86} (\bibinfo {year} {2019})}\BibitemShut {NoStop}%
\bibitem [{\citenamefont {Fiorini}\ and\ \citenamefont
  {Niinikoski}(1984)}]{Fiorini:1983yj}%
  \BibitemOpen
  \bibfield  {author} {\bibinfo {author} {\bibfnamefont {E.}~\bibnamefont
  {Fiorini}}\ and\ \bibinfo {author} {\bibfnamefont {T.}~\bibnamefont
  {Niinikoski}},\ }\href {\doibase 10.1016/0167-5087(84)90449-6} {\bibfield
  {journal} {\bibinfo  {journal} {Nucl. Instrum. Meth.}\ }\textbf {\bibinfo
  {volume} {224}},\ \bibinfo {pages} {83 } (\bibinfo {year}
  {1984})}\BibitemShut {NoStop}%
\bibitem [{\citenamefont {Enss}\ and\ \citenamefont
  {McCammon}(2008)}]{Enss:2008ek}%
  \BibitemOpen
  \bibfield  {author} {\bibinfo {author} {\bibfnamefont {C.}~\bibnamefont
  {Enss}}\ and\ \bibinfo {author} {\bibfnamefont {D.}~\bibnamefont
  {McCammon}},\ }\href {\doibase 10.1007/s10909-007-9611-7} {\bibfield
  {journal} {\bibinfo  {journal} {J. Low Temp. Phys.}\ }\textbf {\bibinfo
  {volume} {151}},\ \bibinfo {pages} {5} (\bibinfo {year} {2008})}\BibitemShut
  {NoStop}%
\bibitem [{\citenamefont {Arnaboldi}\ \emph {et~al.}(2010)\citenamefont
  {Arnaboldi} \emph {et~al.}}]{CUORE-crystals}%
  \BibitemOpen
  \bibfield  {author} {\bibinfo {author} {\bibfnamefont {C.}~\bibnamefont
  {Arnaboldi}} \emph {et~al.},\ }\href {\doibase
  10.1016/j.jcrysgro.2010.06.034} {\bibfield  {journal} {\bibinfo  {journal}
  {J. Cryst. Growth}\ }\textbf {\bibinfo {volume} {312}},\ \bibinfo {pages}
  {2999} (\bibinfo {year} {2010})}\BibitemShut {NoStop}%
\bibitem [{\citenamefont {Redshaw}\ \emph {et~al.}(2009)\citenamefont {Redshaw}
  \emph {et~al.}}]{Redshaw:2009cf}%
  \BibitemOpen
  \bibfield  {author} {\bibinfo {author} {\bibfnamefont {M.}~\bibnamefont
  {Redshaw}} \emph {et~al.},\ }\href {\doibase 10.1103/PhysRevLett.102.212502}
  {\bibfield  {journal} {\bibinfo  {journal} {Phys. Rev. Lett.}\ }\textbf
  {\bibinfo {volume} {102}},\ \bibinfo {pages} {212502} (\bibinfo {year}
  {2009})}\BibitemShut {NoStop}%
\bibitem [{\citenamefont {Scielzo}\ \emph {et~al.}(2009)\citenamefont {Scielzo}
  \emph {et~al.}}]{Scielzo:2009co}%
  \BibitemOpen
  \bibfield  {author} {\bibinfo {author} {\bibfnamefont {N.~D.}\ \bibnamefont
  {Scielzo}} \emph {et~al.},\ }\href {\doibase 10.1103/PhysRevC.80.025501}
  {\bibfield  {journal} {\bibinfo  {journal} {Phys. Rev. C}\ }\textbf {\bibinfo
  {volume} {80}},\ \bibinfo {pages} {025501} (\bibinfo {year}
  {2009})}\BibitemShut {NoStop}%
\bibitem [{\citenamefont {Rahaman}\ \emph {et~al.}(2011)\citenamefont {Rahaman}
  \emph {et~al.}}]{Rahaman:2011wt}%
  \BibitemOpen
  \bibfield  {author} {\bibinfo {author} {\bibfnamefont {S.}~\bibnamefont
  {Rahaman}} \emph {et~al.},\ }\href {\doibase 10.1016/j.physletb.2011.07.078}
  {\bibfield  {journal} {\bibinfo  {journal} {Phys. Lett. B}\ }\textbf
  {\bibinfo {volume} {703}},\ \bibinfo {pages} {412} (\bibinfo {year}
  {2011})}\BibitemShut {NoStop}%
\bibitem [{\citenamefont {{Fehr}}\ \emph {et~al.}(2004)\citenamefont {{Fehr}},
  \citenamefont {{Rehk{\"a}mper}},\ and\ \citenamefont
  {{Halliday}}}]{Fehr:2004jx}%
  \BibitemOpen
  \bibfield  {author} {\bibinfo {author} {\bibfnamefont {M.~A.}\ \bibnamefont
  {{Fehr}}}, \bibinfo {author} {\bibfnamefont {M.}~\bibnamefont
  {{Rehk{\"a}mper}}}, \ and\ \bibinfo {author} {\bibfnamefont {A.~N.}\
  \bibnamefont {{Halliday}}},\ }\href {\doibase 10.1016/j.ijms.2003.11.006}
  {\bibfield  {journal} {\bibinfo  {journal} {International Journal of Mass
  Spectrometry}\ }\textbf {\bibinfo {volume} {232}},\ \bibinfo {pages} {83}
  (\bibinfo {year} {2004})}\BibitemShut {NoStop}%
\bibitem [{\citenamefont {Alduino}\ \emph
  {et~al.}(2018{\natexlab{a}})\citenamefont {Alduino} \emph
  {et~al.}}]{CUORE-rare-processes}%
  \BibitemOpen
  \bibfield  {author} {\bibinfo {author} {\bibfnamefont {C.}~\bibnamefont
  {Alduino}} \emph {et~al.} (\bibinfo {collaboration} {CUORE}),\ }\href
  {\doibase 10.1142/S0217751X18430029} {\bibfield  {journal} {\bibinfo
  {journal} {Int. J. Mod. Phys.}\ }\textbf {\bibinfo {volume} {A33}},\ \bibinfo
  {pages} {1843002} (\bibinfo {year} {2018}{\natexlab{a}})}\BibitemShut
  {NoStop}%
\bibitem [{\citenamefont {Brofferio}\ and\ \citenamefont
  {Dell'Oro}(2018)}]{bolometer-saga}%
  \BibitemOpen
  \bibfield  {author} {\bibinfo {author} {\bibfnamefont {C.}~\bibnamefont
  {Brofferio}}\ and\ \bibinfo {author} {\bibfnamefont {S.}~\bibnamefont
  {Dell'Oro}},\ }\href {\doibase 10.1063/1.5031485} {\bibfield  {journal}
  {\bibinfo  {journal} {Rev. Sci. Instrum.}\ }\textbf {\bibinfo {volume}
  {89}},\ \bibinfo {pages} {121502} (\bibinfo {year} {2018})}\BibitemShut
  {NoStop}%
\bibitem [{\citenamefont {Alduino}\ \emph
  {et~al.}(2016{\natexlab{a}})\citenamefont {Alduino} \emph
  {et~al.}}]{CUORE-0-detector}%
  \BibitemOpen
  \bibfield  {author} {\bibinfo {author} {\bibfnamefont {C.}~\bibnamefont
  {Alduino}} \emph {et~al.} (\bibinfo {collaboration} {CUORE}),\ }\href
  {\doibase 10.1088/1748-0221/11/07/P07009} {\bibfield  {journal} {\bibinfo
  {journal} {JINST}\ }\textbf {\bibinfo {volume} {11}},\ \bibinfo {pages}
  {P07009} (\bibinfo {year} {2016}{\natexlab{a}})}\BibitemShut {NoStop}%
\bibitem [{\citenamefont {Andreotti}\ \emph {et~al.}(2011)\citenamefont
  {Andreotti} \emph {et~al.}}]{cuoricino-final}%
  \BibitemOpen
  \bibfield  {author} {\bibinfo {author} {\bibfnamefont {E.}~\bibnamefont
  {Andreotti}} \emph {et~al.},\ }\href {\doibase
  10.1016/j.astropartphys.2011.02.002} {\bibfield  {journal} {\bibinfo
  {journal} {Astropart. Phys.}\ }\textbf {\bibinfo {volume} {34}},\ \bibinfo
  {pages} {822} (\bibinfo {year} {2011})}\BibitemShut {NoStop}%
\bibitem [{\citenamefont {Alfonso}\ \emph {et~al.}(2015)\citenamefont {Alfonso}
  \emph {et~al.}}]{CUORE-0-results}%
  \BibitemOpen
  \bibfield  {author} {\bibinfo {author} {\bibfnamefont {K.}~\bibnamefont
  {Alfonso}} \emph {et~al.} (\bibinfo {collaboration} {CUORE}),\ }\href
  {\doibase 10.1103/PhysRevLett.115.102502} {\bibfield  {journal} {\bibinfo
  {journal} {Phys. Rev. Lett.}\ }\textbf {\bibinfo {volume} {115}},\ \bibinfo
  {pages} {102502} (\bibinfo {year} {2015})}\BibitemShut {NoStop}%
\bibitem [{\citenamefont {Haller}\ \emph {et~al.}(1984)\citenamefont {Haller}
  \emph {et~al.}}]{Haller1984}%
  \BibitemOpen
  \bibfield  {author} {\bibinfo {author} {\bibfnamefont {E.~E.}\ \bibnamefont
  {Haller}} \emph {et~al.},\ }in\ \href {\doibase 10.1007/978-1-4613-2695-3_2}
  {\emph {\bibinfo {booktitle} {Neutron Transmutation Doping of Semiconductor
  Materials}}},\ \bibinfo {editor} {edited by\ \bibinfo {editor} {\bibfnamefont
  {R.~D.}\ \bibnamefont {Larrabee}}}\ (\bibinfo  {publisher} {Springer US},\
  \bibinfo {address} {Boston, MA},\ \bibinfo {year} {1984})\ pp.\ \bibinfo
  {pages} {21--36}\BibitemShut {NoStop}%
\bibitem [{\citenamefont {Alessandrello}\ \emph {et~al.}(1998)\citenamefont
  {Alessandrello} \emph {et~al.}}]{Alessandrello:1998bf}%
  \BibitemOpen
  \bibfield  {author} {\bibinfo {author} {\bibfnamefont {A.}~\bibnamefont
  {Alessandrello}} \emph {et~al.},\ }\href {\doibase
  10.1016/S0168-9002(98)00458-6} {\bibfield  {journal} {\bibinfo  {journal}
  {Nucl. Instrum. Meth.}\ }\textbf {\bibinfo {volume} {A412}},\ \bibinfo
  {pages} {454} (\bibinfo {year} {1998})}\BibitemShut {NoStop}%
\bibitem [{\citenamefont {Andreotti}\ \emph {et~al.}(2012)\citenamefont
  {Andreotti} \emph {et~al.}}]{CUORE-heaters}%
  \BibitemOpen
  \bibfield  {author} {\bibinfo {author} {\bibfnamefont {E.}~\bibnamefont
  {Andreotti}} \emph {et~al.},\ }\href {\doibase 10.1016/j.nima.2011.10.065}
  {\bibfield  {journal} {\bibinfo  {journal} {Nucl. Instrum. Meth.}\ }\textbf
  {\bibinfo {volume} {A664}},\ \bibinfo {pages} {161} (\bibinfo {year}
  {2012})}\BibitemShut {NoStop}%
\bibitem [{\citenamefont {Alduino}\ \emph {et~al.}(2019)\citenamefont {Alduino}
  \emph {et~al.}}]{CUORE-cryostat}%
  \BibitemOpen
  \bibfield  {author} {\bibinfo {author} {\bibfnamefont {C.}~\bibnamefont
  {Alduino}} \emph {et~al.},\ }\href {\doibase
  10.1016/j.cryogenics.2019.06.011} {\bibfield  {journal} {\bibinfo  {journal}
  {Cryogenics}\ }\textbf {\bibinfo {volume} {102}},\ \bibinfo {pages} {9}
  (\bibinfo {year} {2019})}\BibitemShut {NoStop}%
\bibitem [{\citenamefont {Alessandria}\ \emph {et~al.}(2013)\citenamefont
  {Alessandria} \emph {et~al.}}]{CUORE-4K}%
  \BibitemOpen
  \bibfield  {author} {\bibinfo {author} {\bibfnamefont {F.}~\bibnamefont
  {Alessandria}} \emph {et~al.},\ }\href {\doibase 10.1016/j.nima.2013.06.015}
  {\bibfield  {journal} {\bibinfo  {journal} {Nucl. Instrum. Meth.}\ }\textbf
  {\bibinfo {volume} {A727}},\ \bibinfo {pages} {65} (\bibinfo {year}
  {2013})}\BibitemShut {NoStop}%
\bibitem [{\citenamefont {Buccheri}\ \emph {et~al.}(2014)\citenamefont
  {Buccheri} \emph {et~al.}}]{CUORE-assembly}%
  \BibitemOpen
  \bibfield  {author} {\bibinfo {author} {\bibfnamefont {E.}~\bibnamefont
  {Buccheri}} \emph {et~al.},\ }\href {\doibase 10.1016/j.nima.2014.09.046}
  {\bibfield  {journal} {\bibinfo  {journal} {Nucl. Instrum. Meth. A}\ }\textbf
  {\bibinfo {volume} {768}},\ \bibinfo {pages} {130} (\bibinfo {year}
  {2014})}\BibitemShut {NoStop}%
\bibitem [{\citenamefont {Arnaboldi}\ \emph {et~al.}(2018)\citenamefont
  {Arnaboldi} \emph {et~al.}}]{CUORE-front-end}%
  \BibitemOpen
  \bibfield  {author} {\bibinfo {author} {\bibfnamefont {C.}~\bibnamefont
  {Arnaboldi}} \emph {et~al.},\ }\href {\doibase
  10.1088/1748-0221/13/02/P02026} {\bibfield  {journal} {\bibinfo  {journal}
  {JINST}\ }\textbf {\bibinfo {volume} {13}},\ \bibinfo {pages} {P02026}
  (\bibinfo {year} {2018})}\BibitemShut {NoStop}%
\bibitem [{\citenamefont {Di~Domizio}\ \emph {et~al.}(2018)\citenamefont
  {Di~Domizio} \emph {et~al.}}]{CUORE-DAQ}%
  \BibitemOpen
  \bibfield  {author} {\bibinfo {author} {\bibfnamefont {S.}~\bibnamefont
  {Di~Domizio}} \emph {et~al.},\ }\href {\doibase
  10.1088/1748-0221/13/12/P12003} {\bibfield  {journal} {\bibinfo  {journal}
  {JINST}\ }\textbf {\bibinfo {volume} {13}},\ \bibinfo {pages} {P12003}
  (\bibinfo {year} {2018})}\BibitemShut {NoStop}%
\bibitem [{\citenamefont {Benato}\ \emph {et~al.}(2018)\citenamefont {Benato}
  \emph {et~al.}}]{CUORE-radon-box}%
  \BibitemOpen
  \bibfield  {author} {\bibinfo {author} {\bibfnamefont {G.}~\bibnamefont
  {Benato}} \emph {et~al.},\ }\href {\doibase 10.1088/1748-0221/13/01/P01010}
  {\bibfield  {journal} {\bibinfo  {journal} {JINST}\ }\textbf {\bibinfo
  {volume} {13}},\ \bibinfo {pages} {P01010} (\bibinfo {year}
  {2018})}\BibitemShut {NoStop}%
\bibitem [{\citenamefont {D'Addabbo}\ \emph {et~al.}(2018)\citenamefont
  {D'Addabbo} \emph {et~al.}}]{CUORE-PT}%
  \BibitemOpen
  \bibfield  {author} {\bibinfo {author} {\bibfnamefont {A.}~\bibnamefont
  {D'Addabbo}} \emph {et~al.},\ }\href {\doibase
  10.1016/j.cryogenics.2018.05.001} {\bibfield  {journal} {\bibinfo  {journal}
  {Cryogenics}\ }\textbf {\bibinfo {volume} {93}},\ \bibinfo {pages} {56}
  (\bibinfo {year} {2018})}\BibitemShut {NoStop}%
\bibitem [{\citenamefont {Alduino}\ \emph
  {et~al.}(2018{\natexlab{b}})\citenamefont {Alduino} \emph
  {et~al.}}]{CUORE-PRL2017}%
  \BibitemOpen
  \bibfield  {author} {\bibinfo {author} {\bibfnamefont {C.}~\bibnamefont
  {Alduino}} \emph {et~al.} (\bibinfo {collaboration} {CUORE}),\ }\href
  {\doibase 10.1103/PhysRevLett.120.132501} {\bibfield  {journal} {\bibinfo
  {journal} {Phys. Rev. Lett.}\ }\textbf {\bibinfo {volume} {120}},\ \bibinfo
  {pages} {132501} (\bibinfo {year} {2018}{\natexlab{b}})}\BibitemShut
  {NoStop}%
\bibitem [{\citenamefont {Cushman}\ \emph {et~al.}(2017)\citenamefont {Cushman}
  \emph {et~al.}}]{CUORE-dcs}%
  \BibitemOpen
  \bibfield  {author} {\bibinfo {author} {\bibfnamefont {J.~S.}\ \bibnamefont
  {Cushman}} \emph {et~al.},\ }\href {\doibase 10.1016/j.nima.2016.11.020}
  {\bibfield  {journal} {\bibinfo  {journal} {Nucl. Instrum. Meth. A}\ }\textbf
  {\bibinfo {volume} {844}},\ \bibinfo {pages} {32} (\bibinfo {year}
  {2017})}\BibitemShut {NoStop}%
\bibitem [{\citenamefont {Gatti}\ and\ \citenamefont
  {Manfredi}(1986)}]{Gatti-OT}%
  \BibitemOpen
  \bibfield  {author} {\bibinfo {author} {\bibfnamefont {E.}~\bibnamefont
  {Gatti}}\ and\ \bibinfo {author} {\bibfnamefont {P.~F.}\ \bibnamefont
  {Manfredi}},\ }\href {\doibase 10.1007/BF02822156} {\bibfield  {journal}
  {\bibinfo  {journal} {Riv. Nuovo Cim.}\ }\textbf {\bibinfo {volume} {9N1}},\
  \bibinfo {pages} {1} (\bibinfo {year} {1986})}\BibitemShut {NoStop}%
\bibitem [{\citenamefont {Di~Domizio}\ \emph {et~al.}(2011)\citenamefont
  {Di~Domizio}, \citenamefont {Orio},\ and\ \citenamefont
  {Vignati}}]{CUORE-0-opttrigger}%
  \BibitemOpen
  \bibfield  {author} {\bibinfo {author} {\bibfnamefont {S.}~\bibnamefont
  {Di~Domizio}}, \bibinfo {author} {\bibfnamefont {F.}~\bibnamefont {Orio}}, \
  and\ \bibinfo {author} {\bibfnamefont {M.}~\bibnamefont {Vignati}},\ }\href
  {\doibase 10.1088/1748-0221/6/02/P02007} {\bibfield  {journal} {\bibinfo
  {journal} {JINST}\ }\textbf {\bibinfo {volume} {6}},\ \bibinfo {pages}
  {P02007} (\bibinfo {year} {2011})}\BibitemShut {NoStop}%
\bibitem [{\citenamefont {Alfonso}\ \emph {et~al.}(2018)\citenamefont {Alfonso}
  \emph {et~al.}}]{CUORE-pulser}%
  \BibitemOpen
  \bibfield  {author} {\bibinfo {author} {\bibfnamefont {K.}~\bibnamefont
  {Alfonso}} \emph {et~al.},\ }\href {\doibase 10.1088/1748-0221/13/02/P02029}
  {\bibfield  {journal} {\bibinfo  {journal} {JINST}\ }\textbf {\bibinfo
  {volume} {13}},\ \bibinfo {pages} {P02029} (\bibinfo {year}
  {2018})}\BibitemShut {NoStop}%
\bibitem [{\citenamefont {Alduino}\ \emph
  {et~al.}(2016{\natexlab{b}})\citenamefont {Alduino} \emph
  {et~al.}}]{CUORE-0-analysis-techniques}%
  \BibitemOpen
  \bibfield  {author} {\bibinfo {author} {\bibfnamefont {C.}~\bibnamefont
  {Alduino}} \emph {et~al.} (\bibinfo {collaboration} {CUORE}),\ }\href
  {\doibase 10.1103/PhysRevC.93.045503} {\bibfield  {journal} {\bibinfo
  {journal} {Phys. Rev. C}\ }\textbf {\bibinfo {volume} {93}},\ \bibinfo
  {pages} {045503} (\bibinfo {year} {2016}{\natexlab{b}})}\BibitemShut
  {NoStop}%
\bibitem [{\citenamefont {Mahalanobis}(1936)}]{Mahalanobis:1936tj}%
  \BibitemOpen
  \bibfield  {author} {\bibinfo {author} {\bibfnamefont {P.~C.}\ \bibnamefont
  {Mahalanobis}},\ }\href@noop {} {\bibfield  {journal} {\bibinfo  {journal}
  {Proc. Natl. Inst. Sci. India}\ ,\ \bibinfo {pages} {49}} (\bibinfo {year}
  {1936})}\BibitemShut {NoStop}%
\bibitem [{\citenamefont {Drobizhev}(2018)}]{alexey}%
  \BibitemOpen
  \bibfield  {author} {\bibinfo {author} {\bibfnamefont {A.}~\bibnamefont
  {Drobizhev}},\ }\emph {\bibinfo {title} {{Searching for the $0\nu\beta\beta$
  decay of $^{130}Te$ with the ton-scale CUORE bolometer array}}},\ \href@noop
  {} {Ph.D. thesis},\ \bibinfo  {school} {{University of California, Berkeley}}
  (\bibinfo {year} {2018})\BibitemShut {NoStop}%
\bibitem [{\citenamefont {Marini}(2018)}]{laura}%
  \BibitemOpen
  \bibfield  {author} {\bibinfo {author} {\bibfnamefont {L.}~\bibnamefont
  {Marini}},\ }\emph {\bibinfo {title} {{The CUORE experiment: from the
  commissioning to the first $0\nu\beta\beta$ limit}}},\ \href@noop {} {Ph.D.
  thesis},\ \bibinfo  {school} {{Universit\`a degli Studi di Genova}} (\bibinfo
  {year} {2018})\BibitemShut {NoStop}%
\bibitem [{\citenamefont {Alduino}\ \emph
  {et~al.}(2017{\natexlab{a}})\citenamefont {Alduino} \emph
  {et~al.}}]{CUORE-projected-background}%
  \BibitemOpen
  \bibfield  {author} {\bibinfo {author} {\bibfnamefont {C.}~\bibnamefont
  {Alduino}} \emph {et~al.} (\bibinfo {collaboration} {CUORE}),\ }\href
  {\doibase 10.1140/epjc/s10052-017-5080-6} {\bibfield  {journal} {\bibinfo
  {journal} {Eur. Phys. J.}\ }\textbf {\bibinfo {volume} {C77}},\ \bibinfo
  {pages} {543} (\bibinfo {year} {2017}{\natexlab{a}})}\BibitemShut {NoStop}%
\bibitem [{\citenamefont {Alduino}\ \emph
  {et~al.}(2017{\natexlab{b}})\citenamefont {Alduino} \emph
  {et~al.}}]{CUORE-0-bkgmodel}%
  \BibitemOpen
  \bibfield  {author} {\bibinfo {author} {\bibfnamefont {C.}~\bibnamefont
  {Alduino}} \emph {et~al.} (\bibinfo {collaboration} {CUORE}),\ }\href
  {\doibase 10.1140/epjc/s10052-016-4498-6} {\bibfield  {journal} {\bibinfo
  {journal} {Eur. Phys. J.}\ }\textbf {\bibinfo {volume} {C77}},\ \bibinfo
  {pages} {13} (\bibinfo {year} {2017}{\natexlab{b}})}\BibitemShut {NoStop}%
\bibitem [{\citenamefont {Caldwell}\ \emph {et~al.}(2009)\citenamefont
  {Caldwell}, \citenamefont {Koll\'{a}r},\ and\ \citenamefont
  {Kr\"{o}ninger}}]{bat}%
  \BibitemOpen
  \bibfield  {author} {\bibinfo {author} {\bibfnamefont {A.}~\bibnamefont
  {Caldwell}}, \bibinfo {author} {\bibfnamefont {D.}~\bibnamefont
  {Koll\'{a}r}}, \ and\ \bibinfo {author} {\bibfnamefont {K.}~\bibnamefont
  {Kr\"{o}ninger}},\ }\href {\doibase 10.1016/j.cpc.2009.06.026} {\bibfield
  {journal} {\bibinfo  {journal} {Computer Physics Communications}\ }\textbf
  {\bibinfo {volume} {180}},\ \bibinfo {pages} {2197–2209} (\bibinfo {year}
  {2009})}\BibitemShut {NoStop}%
\bibitem [{\citenamefont {Kotila}\ and\ \citenamefont
  {Iachello}(2012)}]{Kotila:2012zza}%
  \BibitemOpen
  \bibfield  {author} {\bibinfo {author} {\bibfnamefont {J.}~\bibnamefont
  {Kotila}}\ and\ \bibinfo {author} {\bibfnamefont {F.}~\bibnamefont
  {Iachello}},\ }\href {\doibase 10.1103/PhysRevC.85.034316} {\bibfield
  {journal} {\bibinfo  {journal} {Phys. Rev.}\ }\textbf {\bibinfo {volume}
  {C85}},\ \bibinfo {pages} {034316} (\bibinfo {year} {2012})}\BibitemShut
  {NoStop}%
\bibitem [{\citenamefont {Engel}\ and\ \citenamefont
  {Men\'endez}(2017)}]{Engel:2016xgb}%
  \BibitemOpen
  \bibfield  {author} {\bibinfo {author} {\bibfnamefont {J.}~\bibnamefont
  {Engel}}\ and\ \bibinfo {author} {\bibfnamefont {J.}~\bibnamefont
  {Men\'endez}},\ }\href {\doibase 10.1088/1361-6633/aa5bc5} {\bibfield
  {journal} {\bibinfo  {journal} {Rept. Prog. Phys.}\ }\textbf {\bibinfo
  {volume} {80}},\ \bibinfo {pages} {046301} (\bibinfo {year}
  {2017})}\BibitemShut {NoStop}%
\bibitem [{\citenamefont {Barea}\ \emph {et~al.}(2015)\citenamefont {Barea},
  \citenamefont {Kotila},\ and\ \citenamefont {Iachello}}]{Barea:2015kwa}%
  \BibitemOpen
  \bibfield  {author} {\bibinfo {author} {\bibfnamefont {J.}~\bibnamefont
  {Barea}}, \bibinfo {author} {\bibfnamefont {J.}~\bibnamefont {Kotila}}, \
  and\ \bibinfo {author} {\bibfnamefont {F.}~\bibnamefont {Iachello}},\ }\href
  {\doibase 10.1103/PhysRevC.91.034304} {\bibfield  {journal} {\bibinfo
  {journal} {Phys. Rev.}\ }\textbf {\bibinfo {volume} {C91}},\ \bibinfo {pages}
  {034304} (\bibinfo {year} {2015})}\BibitemShut {NoStop}%
\bibitem [{\citenamefont {\v{S}imkovic}\ \emph {et~al.}(2013)\citenamefont
  {\v{S}imkovic} \emph {et~al.}}]{Simkovic:2013qiy}%
  \BibitemOpen
  \bibfield  {author} {\bibinfo {author} {\bibfnamefont {F.}~\bibnamefont
  {\v{S}imkovic}} \emph {et~al.},\ }\href {\doibase 10.1103/PhysRevC.87.045501}
  {\bibfield  {journal} {\bibinfo  {journal} {Phys. Rev.}\ }\textbf {\bibinfo
  {volume} {C87}},\ \bibinfo {pages} {045501} (\bibinfo {year}
  {2013})}\BibitemShut {NoStop}%
\bibitem [{\citenamefont {Hyv{\"a}rinen}\ and\ \citenamefont
  {Suhonen}(2015)}]{Hyvarinen:2015bda}%
  \BibitemOpen
  \bibfield  {author} {\bibinfo {author} {\bibfnamefont {J.}~\bibnamefont
  {Hyv{\"a}rinen}}\ and\ \bibinfo {author} {\bibfnamefont {J.}~\bibnamefont
  {Suhonen}},\ }\href {\doibase 10.1103/PhysRevC.91.024613} {\bibfield
  {journal} {\bibinfo  {journal} {Phys. Rev.}\ }\textbf {\bibinfo {volume}
  {C91}},\ \bibinfo {pages} {024613} (\bibinfo {year} {2015})}\BibitemShut
  {NoStop}%
\bibitem [{\citenamefont {Men\'endez}\ \emph {et~al.}(2009)\citenamefont
  {Men\'endez} \emph {et~al.}}]{Menendez:2008jp}%
  \BibitemOpen
  \bibfield  {author} {\bibinfo {author} {\bibfnamefont {J.}~\bibnamefont
  {Men\'endez}} \emph {et~al.},\ }\href {\doibase
  10.1016/j.nuclphysa.2008.12.005} {\bibfield  {journal} {\bibinfo  {journal}
  {Nucl. Phys.}\ }\textbf {\bibinfo {volume} {A818}},\ \bibinfo {pages} {139}
  (\bibinfo {year} {2009})}\BibitemShut {NoStop}%
\bibitem [{\citenamefont {Rodriguez}\ and\ \citenamefont
  {Martinez-Pinedo}(2010)}]{Rodriguez:2010mn}%
  \BibitemOpen
  \bibfield  {author} {\bibinfo {author} {\bibfnamefont {T.~R.}\ \bibnamefont
  {Rodriguez}}\ and\ \bibinfo {author} {\bibfnamefont {G.}~\bibnamefont
  {Martinez-Pinedo}},\ }\href {\doibase 10.1103/PhysRevLett.105.252503}
  {\bibfield  {journal} {\bibinfo  {journal} {Phys. Rev. Lett.}\ }\textbf
  {\bibinfo {volume} {105}},\ \bibinfo {pages} {252503} (\bibinfo {year}
  {2010})}\BibitemShut {NoStop}%
\bibitem [{\citenamefont {L\'opez~Vaquero}\ \emph {et~al.}(2013)\citenamefont
  {L\'opez~Vaquero}, \citenamefont {Rodr\'iguez},\ and\ \citenamefont
  {Egido}}]{Vaquero:2014dna}%
  \BibitemOpen
  \bibfield  {author} {\bibinfo {author} {\bibfnamefont {N.}~\bibnamefont
  {L\'opez~Vaquero}}, \bibinfo {author} {\bibfnamefont {T.~R.}\ \bibnamefont
  {Rodr\'iguez}}, \ and\ \bibinfo {author} {\bibfnamefont {J.~L.}\ \bibnamefont
  {Egido}},\ }\href {\doibase 10.1103/PhysRevLett.111.142501} {\bibfield
  {journal} {\bibinfo  {journal} {Phys. Rev. Lett.}\ }\textbf {\bibinfo
  {volume} {111}},\ \bibinfo {pages} {142501} (\bibinfo {year}
  {2013})}\BibitemShut {NoStop}%
\bibitem [{\citenamefont {Yao}\ \emph {et~al.}(2015)\citenamefont {Yao} \emph
  {et~al.}}]{Yao:2014uta}%
  \BibitemOpen
  \bibfield  {author} {\bibinfo {author} {\bibfnamefont {J.~M.}\ \bibnamefont
  {Yao}} \emph {et~al.},\ }\href {\doibase 10.1103/PhysRevC.91.024316}
  {\bibfield  {journal} {\bibinfo  {journal} {Phys. Rev.}\ }\textbf {\bibinfo
  {volume} {C91}},\ \bibinfo {pages} {024316} (\bibinfo {year}
  {2015})}\BibitemShut {NoStop}%
\bibitem [{\citenamefont {Mustonen}\ and\ \citenamefont
  {Engel}(2013)}]{Mustonen:2013zu}%
  \BibitemOpen
  \bibfield  {author} {\bibinfo {author} {\bibfnamefont {M.~T.}\ \bibnamefont
  {Mustonen}}\ and\ \bibinfo {author} {\bibfnamefont {J.}~\bibnamefont
  {Engel}},\ }\href {\doibase 10.1103/PhysRevC.87.064302} {\bibfield  {journal}
  {\bibinfo  {journal} {Phys. Rev.}\ }\textbf {\bibinfo {volume} {C87}},\
  \bibinfo {pages} {064302} (\bibinfo {year} {2013})}\BibitemShut {NoStop}%
\bibitem [{\citenamefont {Neacsu}\ and\ \citenamefont
  {Horoi}(2015)}]{Neacsu:2014bia}%
  \BibitemOpen
  \bibfield  {author} {\bibinfo {author} {\bibfnamefont {A.}~\bibnamefont
  {Neacsu}}\ and\ \bibinfo {author} {\bibfnamefont {M.}~\bibnamefont {Horoi}},\
  }\href {\doibase 10.1103/PhysRevC.91.024309} {\bibfield  {journal} {\bibinfo
  {journal} {Phys. Rev.}\ }\textbf {\bibinfo {volume} {C91}},\ \bibinfo {pages}
  {024309} (\bibinfo {year} {2015})}\BibitemShut {NoStop}%
\bibitem [{\citenamefont {Meroni}\ \emph {et~al.}(2013)\citenamefont {Meroni},
  \citenamefont {Petcov},\ and\ \citenamefont {Simkovic}}]{Meroni:2012qf}%
  \BibitemOpen
  \bibfield  {author} {\bibinfo {author} {\bibfnamefont {A.}~\bibnamefont
  {Meroni}}, \bibinfo {author} {\bibfnamefont {S.~T.}\ \bibnamefont {Petcov}},
  \ and\ \bibinfo {author} {\bibfnamefont {F.}~\bibnamefont {Simkovic}},\
  }\href {\doibase 10.1007/JHEP02(2013)025} {\bibfield  {journal} {\bibinfo
  {journal} {JHEP}\ }\textbf {\bibinfo {volume} {02}},\ \bibinfo {pages} {025}
  (\bibinfo {year} {2013})}\BibitemShut {NoStop}%
\bibitem [{\citenamefont {Armengaud}\ \emph {et~al.}(2019)\citenamefont
  {Armengaud} \emph {et~al.}}]{CUPID-Mo-instrument}%
  \BibitemOpen
  \bibfield  {author} {\bibinfo {author} {\bibfnamefont {E.}~\bibnamefont
  {Armengaud}} \emph {et~al.} (\bibinfo {collaboration} {CUPID-Mo}),\
  }\href@noop {} {\  (\bibinfo {year} {2019})},\ \Eprint
  {http://arxiv.org/abs/1909.02994} {arXiv:1909.02994 [physics.ins-det]}
  \BibitemShut {NoStop}%
\bibitem [{\citenamefont {Armstrong}\ \emph {et~al.}(2019)\citenamefont
  {Armstrong} \emph {et~al.}}]{CUPID-preCDR}%
  \BibitemOpen
  \bibfield  {author} {\bibinfo {author} {\bibfnamefont {W.~R.}\ \bibnamefont
  {Armstrong}} \emph {et~al.} (\bibinfo {collaboration} {CUPID}),\ }\href@noop
  {} {\  (\bibinfo {year} {2019})},\ \Eprint {http://arxiv.org/abs/1907.09376}
  {arXiv:1907.09376 [physics.ins-det]} \BibitemShut {NoStop}%
\end{thebibliography}%

\end{document}